\documentclass{aa}                 
\usepackage{graphics}              
\usepackage{times}
\begin{document}


\title{Microwave plasma emission of a flare on AD Leo}

\author{A.V.~Stepanov\inst{1}
  \and  B.~Kliem\inst{2} 
  \and  V.V.~Zaitsev\inst{3}
  \and  E.~F\"urst\inst{4}  \and A.~Jessner\inst{4}
  \and  A.~Kr\"uger\inst{2} \and J.~Hildebrandt\inst{2}
  \and  J.H.M.M.~Schmitt\inst{5}}

\institute{Pulkovo Observatory, 196140 St.\,Petersburg, Russia; 
           stepanov@gao.spb.ru
     \and  Astrophysikalisches Institut Potsdam, D-14482~Potsdam, Germany; 
           bkliem@aip.de
     \and  Institute of Applied Physics, 603600 Nizhny Novgorod, Russia; 
           za130@appl.sci-nnov.ru
     \and  Max-Planck-Institut f\"ur Radioastronomie, Auf dem H\"ugel 69, 
           D-53121 Bonn, Germany; efuerst@mpifr-bonn.mpg.de
     \and  Universit\"atssternwarte Hamburg, D-21029 Hamburg, Germany; 
           jschmitt@hs.uni-hamburg.de}

\titlerunning{Microwave plasma emission of a flare on AD Leo}
\authorrunning{A.V.~Stepanov et~al.}

\offprints{B.Kliem}

\date{Received 22 September 2000 / Accepted 6 April 2001}

\abstract{
An intense radio flare on the dMe star AD~Leo, observed with the
Effelsberg radio telescope and spectrally resolved in a band of 480~MHz
centred at 4.85~GHz is analysed. 
A lower limit of the brightness temperature of the totally
right handed polarized emission is estimated as
$T_{\rm b}\sim5\times10^{10}$~K (with values 
$T_{\rm b}\ga3\times10^{13}$~K considered to be more probable), 
which requires a coherent radio emission process. 
In the interpretation we favour fundamental plasma radiation by
mildly relativistic electrons trapped in a hot and dense coronal loop
above electron cyclotron maser emission. This leads to densities and
magnetic field strengths in the radio source of
$n\sim2\times10^{11}$~cm$^{-3}$ and $B\sim800$~G. Quasi-periodic
pulsations during the decay phase of the event suggest a loop radius of
$r\sim7\times10^8$~cm. A filamentary corona is implied in which the dense
radio source is embedded in hot thin plasma with temperature
$T\ge2\times10^7$~K 
and density $n_{\rm ext}\le10^{-2}n$. 
Runaway acceleration by sub-Dreicer electric fields in a magnetic
loop is found to supply a sufficient number of energetic electrons. 
}

\maketitle

\section{Introduction}
\label{intro}

Several nearby flare stars occasionally show intense radio outbursts.  The
most prominent events were observed from AD~Leonis, a very active single M
dwarf star (Gliese 388, dM3.5e, $R_\ast=3.5\times10^{10}$~cm, 
$d=4.85~{\rm pc}=1.55\times10^{19}$~cm). Its radio flares have been
detected mainly in the decimetric range at various observing wavelengths
around 20~cm (Lang et al.\ \cite{Lan83}; Lang \& Willson \cite{LW86};
White et al.\ \cite{WKJ86}; Jackson et al.\ \cite{JKW89}; G\"udel et al.\
\cite{Gud89}; Bastian et al.\ \cite{BBDD90}; Abada-Simon et al.\
\cite{Aba94a, Aba97}). The estimates of the brightness temperature of
these emissions range from $T_{\rm b}\ga10^{10}$~K to 
$T_{\rm b}\sim10^{16}$~K, which all imply a coherent emission process. The
circular polarization is often high, but can vary from event to event
between negligible values and $\approx100$\,\%. The emission can be
different or even absent at a neighbouring frequency only $\sim10$\,\%
apart. On several occasions, dynamic spectra have shown a wealth of fine
structures in the time-frequency plane (pulsations, sudden reductions,
spikes, fast-drift bursts), largely similar to solar dynamic radio
spectra. All these properties --- high polarization degree, narrow
bandwidth, and fine structures --- require a coherent emission process. 

AD~Leo flare observations in the microwave range, performed mainly near
6~cm, are less numerous and have produced a less coherent picture so far.
Possibly the observed events belong to two or three groups. (1) There are
long-lasting ($t\!\la\!1$~hour), relatively weakly polarized bursts of low
to moderate flux density, $F_{\rm peak}=1.4\mbox{--}20$~mJy (Gary et al.\
\cite{GBB87}; Jackson et al.\ \cite{JKW89}; Rodon\`o et al.\
\cite{Rod90}). If a source radius equal to the stellar radius is assumed,
these flux densities correspond to brightness temperatures 
$T_{\rm b}\sim(0.2\mbox{--}2)\times10^9$~K, consistent with incoherent
emission by mildly relativistic electrons. (2) A more impulsive,
completely left-handed polarized event of $\approx5$~min duration and flux
density $F_{\rm peak}\approx30$~mJy was observed by Gary et al.\
(\cite{GBB87}). Of similar nature may be the impulsive onset phase of the
gradual burst observed by Rodon\`o et al.\ (\cite{Rod90}), which had
$\approx2$~min duration, a flux density $F_{\rm peak}\approx30$~mJy, and
was seen only at 2~cm, not at 6 or 20~cm (no information on the
polarization was provided). If the source diameter is estimated from the
rise time using the Alfv\'en velocity 
$v_{\rm A}\approx7\times10^8\,\omega_{\rm c}/\omega_{\rm p}$~cm\,s$^{-1}$
(where the ratio of electron cyclotron and plasma frequencies is assumed
here to be $\omega_{\rm c}/\omega_{\rm p}\sim1$ for the purpose of an
estimate), one obtains $T_{\rm b}\sim7\times10^{10}$~K for the burst
observed by Gary et al.\ (\cite{GBB87}) and $T_{\rm b}\sim3\times10^9$~K
for the burst observed by Rodon\`o et al.\ (\cite{Rod90}). At least the
former of these bursts requires a coherent emission mechanism. (3)
Lecacheux et al.\ (\cite{Lec93}) observed one event of $\approx400$~mJy
peak flux and $\sim4$~s duration, which was considered likely to originate
from AD~Leo. The burst was resolved by an acousto-optical spectrograph
operating in the 4.5--5~GHz range at the Arecibo telescope. It was
extremely narrow-banded, clearly discernible only between 4.78 and
4.91~GHz, and only moderately polarized ($\approx20$\,\%). The brightness
temperature was estimated as 
$T_{\rm b}\approx3\times10^{10}(R_\ast/r_{\rm source})^2$~K, where it is
clear from the short duration that the source radius must have been much
smaller than the stellar radius, $r_{\rm source}\ll R_\ast$; again a
coherent emission mechanism is required. 

The stellar origin of these radio bursts has been confirmed, in the
majority of cases, by simultaneous observing with up to three widely
separated telescopes, by imaging with the VLA, by eliminating interference
based on its characteristic appearances in dynamic spectra, or (in case of
single dishes) by using two feeds simultaneously with one feed in an
``off'' position. Also the similarity of the fine structures with solar
radio fine structures suggests a stellar origin. 

In the 2--6~cm range, a transition from coherent emission at long
wavelengths to incoherent emission at short wavelengths appears to occur
in the flare emission of AD~Leo, similar to the situation at other flare
stars (Bastian \cite{Bas90}) and the Sun, perhaps shifted to somewhat
shorter wavelengths at flare stars. It is understood as being due to three
effects. First, the characteristic frequencies of coherent emission
($s\omega_{\rm c}/2\pi;~s\!=\!1,\,2$ for the cyclotron maser and
$s\omega_{\rm p}/2\pi;~s\!=\!1,\,2$ for plasma emission) are limited to
$\sim10$~GHz by the available field strengths and plasma densities in
stellar coronae. Second, a general increase of the ratio 
$\omega_{\rm c}/\omega_{\rm p}$ with decreasing height in the corona,
which makes cyclotron emission processes more important and plasma
emission less important or quenches the latter completely. Third,
increasing free-free absorption of plasma emission and of low-harmonic
cyclotron maser emission at increasing plasma density (implied by higher
observing frequency $\nu$), see, e.g., Dulk (\cite{Dul85}), Benz
(\cite{Ben84}). However, also similar to the solar case, coherent
emissions observed in this range, albeit being rare and possibly becoming
progressively weak at decreasing wavelength, are likely to originate
closest to, or from within, the primary flare energy release volume (e.g.,
Benz et al.\ \cite{BMM01}). 

We have therefore begun a search for bursts from AD~Leo in the 6~cm range
using the Effelsberg radio telescope and a bandwidth of $\approx500$~MHz
centred near 4.85~GHz, which provides a sensitivity limit of $\sim50$~mJy
in dynamic spectra. A well developed, intense burst was detected, which
warrants a detailed study, presented in this paper. The burst had a peak
flux of $\approx300$~mJy, duration of $\ga1$~min, a broadband spectrum
that varied slightly during the event, and was $\approx100$\,\%
right-handed circularly polarized. The estimated brightness temperature, 
$T_{\rm b}>5\times10^{10}$~K, implies a coherent emission process.
The temporal modulations of the decay phase resemble quasi-periodic
pulsations. 

Simultaneous soft X-ray observations using ROSAT were not successful,
since the burst occurred during source occultation by the Earth. 

Our analysis of the event focuses on the nonthermal radio emission
mechanism.
There are two mechanisms that can lead to the required high 
brightness temperature: electron cyclotron maser (ECM) emission and plasma
emission (e.g., Dulk \cite{Dul85}; Kuijpers \cite{Kui89}).
Since gyroresonance absorption of ECM emission by ambient thermal electrons
presents a severe problem for temperatures and densities supposed to be
typical of M dwarf coronae, we consider plasma emission in detail.
The high polarization degree requires that the plasma emission arises at
the fundamental. We assume that a population of mildly relativistic
electrons is trapped in a coronal loop, develops a loss-cone distribution,
and excites electrostatic upper-hybrid modes (analogous to the canonical
source model of broadband solar plasma emission). Calculations of the
brightness temperature of fundamental and harmonic plasma radiation for
typical M dwarf coronal parameters confirm the possibility of fundamental
emission. This mechanism implies smaller magnetic field strengths and
higher plasma densities than ECM emission. Electron acceleration by a
sub-Dreicer DC electric field at the chromospheric footpoints of a coronal
loop is briefly considered and
found to be consistent with the requirements on the number and
energy of accelerated electrons in the observed radio flare. Finally, a
consideration of the pulsations supports the assumption that the radio
source was formed by particle trapping in a loop. 

The observations are presented in Sect.\,\ref{obs}, the ECM and plasma
emission processes are discussed and source parameters are estimated in
Sect.\,\ref{mechanism}, the calculation of the brightness temperature for
fundamental and harmonic plasma emission is given in Sect.\,\ref{bright},
and the particle acceleration is considered in Sect.\,\ref{accel}. The
origin of the pulsations is discussed in Sect.\,\ref{puls}, and the
results are summarized in Sect.\,\ref{concl}.

\section{Radio observations of the AD~Leo flare at Effelsberg}
\label{obs}

The radio observations of AD~Leo at Effelsberg lasted a total of about 27
hours on the days May 17-19, 1997. During the first two days the
meteorological conditions were not favourable so that possible bursts
could not be identified with certainty, but a well developed burst with a
duration of $\ga1$~min was recorded on May 19, 1997, 18.945 UT.

The observations were performed with the new 8192-channel autocorrelation
system (AK) and in parallel with the pulsar receiving equipment. The
centre frequency of the measurements was 4.85~GHz. The total effective
bandwidth was 500~MHz for the pulsar mode and 480~MHz for the
spectroscopic mode. Both observing systems recorded the two total power
signals (LHC and RHC) of the four channel receiving system.

\subsection{The pulsar mode}

Left and right handed circular polarized radio power was detected from the
two feeds of the receiver, one being centred on the source and the other
one in an ``off'' position.  The same centre frequency of 4.85~GHz was
employed and the detectors operated at an effective bandwidth of 500~MHz.
Using fast (10~MHz) V/f converters, the four frequency encoded intensity
channels provided the input to the Effelsberg Pulsar Observing System
(EPOS).  The V/f frequencies of the four channels, typically about 4~MHz,
were continuously measured in 1.0~ms intervals.  Hence a resolution of
$2.5\times10^{-4}$ in detected power and 1.0~ms in time were the limits of
the pulsar data aquisition as it was operated.  For a system temperature
of 30~K and a telescope efficiency of 46\,\%, about 6 counts per sample
correspond to the achievable sensitivity of 32~mJy.  After each collection
of 1024 data samples, these, together with a $\mu$s timestamp,
synchronized to the station clock, were recorded on disk.

\begin{figure}   
   \resizebox{\hsize}{!}{\includegraphics{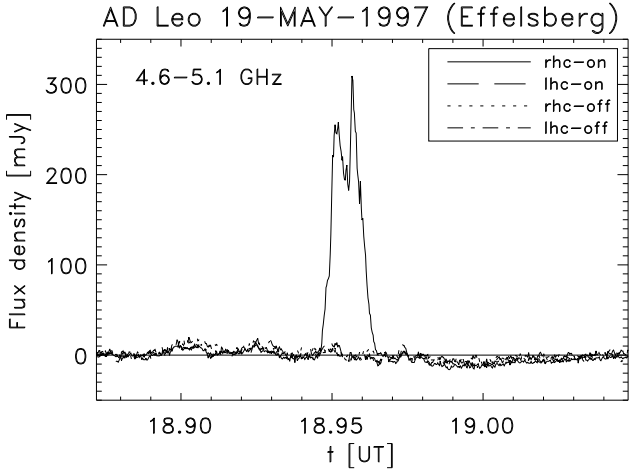}}
\caption[]
{Time profile of the AD~Leo flare observed at the Effelsberg
100~m radiotelescope on May 19, 1997, 18.945 UT at 4.85~GHz. 
Right and left handed circularly polarized flux on-source and off-source
are plotted.}
\label{overview}
\end{figure}

Figure\,\ref{overview} shows the time profile of the event. Here the data
were binned down to a time resolution of 1~s to obtain  a better signal to
noise ratio. The contribution of the ground radiation has been removed by
a parabolic fit to the data. The r.m.s.\ noise after binning is about
2~mJy. Some low level structures, especially before 18.94~UT, are visible
in all four channels and are very likely caused by weather effects. Only
the feed centred on the source observed the strong burst of radio
emission, which was also highly polarized. While the typical flux density
of the radio burst is $\sim200$~mJy in the right handed polarization, the
signal in the left handed polarization channel is below the 3 sigma level
of 6~mJy. The polarization degree of the detected signal is, therefore,
larger than 97\,\%.

\subsection{The spectroscopic mode}

The new AK was used to record the left handed and right handed
polarization with a large bandwidth.  Four frequency bands were recorded,
each 160~MHz wide.  The centre frequencies of the individual bands were
-180~MHz, -60~MHz, +60~MHz, and +180~MHz with respect to 4850~MHz.  Each
band consisted of 128 channels that were 1.25~MHz wide.  The integration
time was set to 128~ms.  The reduction of the data was done by calculating
$(S-B)/B$ (where $S$ is the signal and $B$ is the background flux), and
the calibration used a flux density of 7.5~Jy for 3C286.  Where the
sub-bands overlapped, the 16 overlapping data points (20~MHz) were skipped
because different baseline characteristics at the end of the bands
prohibited their individual comparison.  Differences of the flux density
between the edges of adjacent bands were corrected by applying linear
baselines in such a way that the corrected value became equal to the mean
of the adjacent edges. The four bands were then merged to the effective
bandwidth of 480~MHz.  In most cases the systematic uncertainty resulting
from this procedure is below 30~mJy, but values of up to 100~mJy were
occasionally noted.  It is assumed that there is no other flare emission
outside the flare event visible in Fig.\,\ref{overview}.  A linear
baseline along the time axis was subtracted to remove the background
emission (i.e., the flux level before and after the considered event was
set to zero).  Finally, a method of unsharp masking invented by Sofue \&
Reich (\cite{SR79}) was used to remove ``scanning effects'' along the time
axis.  The r.m.s.\ in quiet regions of the spectral data was found to be
$\approx50$~mJy.

\begin{figure*}   
 \begin{center}
    \resizebox{0.92\hsize}{!}{\includegraphics{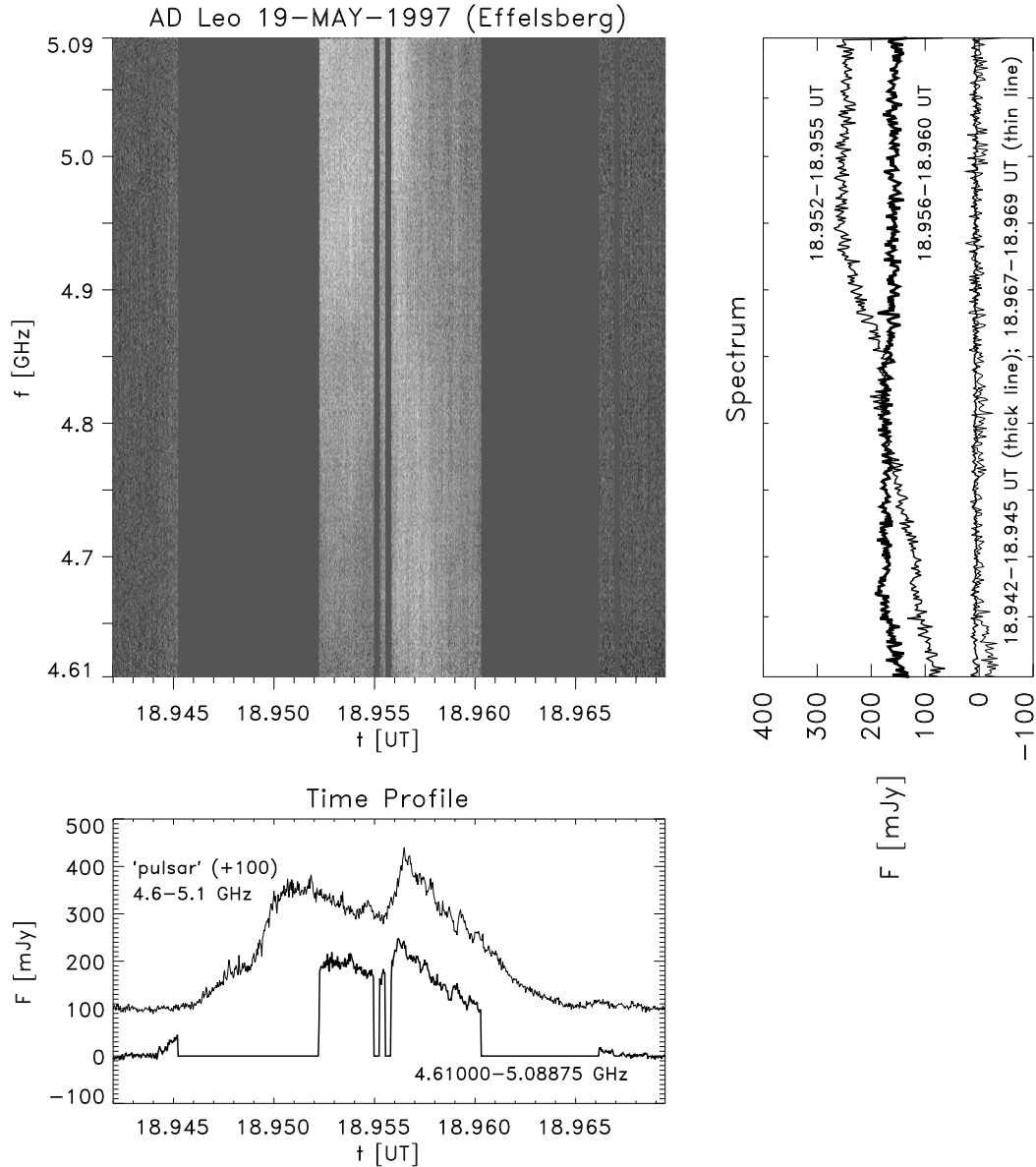}}
 \end{center}
\vspace{-0.9cm}
\caption[]
{Plot of AK90 spectroscopic observations of the AD~Leo flare on May 19,
1997 (top left). Intervals with no reliable spectra were cleared. The
time profile averaged over the spectrum is shown in the bottom panel
together with the pulsar (EPOS) data (binned down to the same time
resolution and offset by 100~mJy for clarity). The top right panel shows
average spectra in the four time intervals indicated.}
\label{radio-spec}
\end{figure*}

The resulting dynamic spectrum is shown in Fig.\,\ref{radio-spec} for the
right handed polarization.  Due to internal problems of the new AK, we
lost the signal during certain periods.  These time intervals have been
cleared in Fig.\,\ref{radio-spec}. No signals were detected in left handed
polarization.  The right handed data show a clear signal between 18.945
and 18.965 universal time.  Between 18.952 and 18.955~UT the largest
differences between adjacent bands were detected (mean $\approx80$~mJy,
peak 100~mJy) at frequencies near 4630~MHz.  The data in this time
interval have, therefore, the lowest accuracy.  The other time slices show
systematic errors of only $\approx30$~mJy in addition to the statistical
error of $\approx50$~mJy.

\subsection{Spectral and temporal details}

The bottom panel of Fig.\,\ref{radio-spec} displays the flux obtained in
spectral mode integrated over the whole bandwidth.  The comparison with
the time profile obtained from the pulsar data set, binned down to the
same time resolution of 125~ms, shows excellent agreement after 18.956~UT,
but also some deviations in the time interval 18.952--18.955~UT.  These
deviations reflect the systematic errors discussed above.  The upper right
panel shows the average spectra in each of the four time intervals
indicated.  Some spectral evolution during the lifetime of the burst is
evident.

\begin{figure}[t]   
    \resizebox{\hsize}{!}                
              {\includegraphics{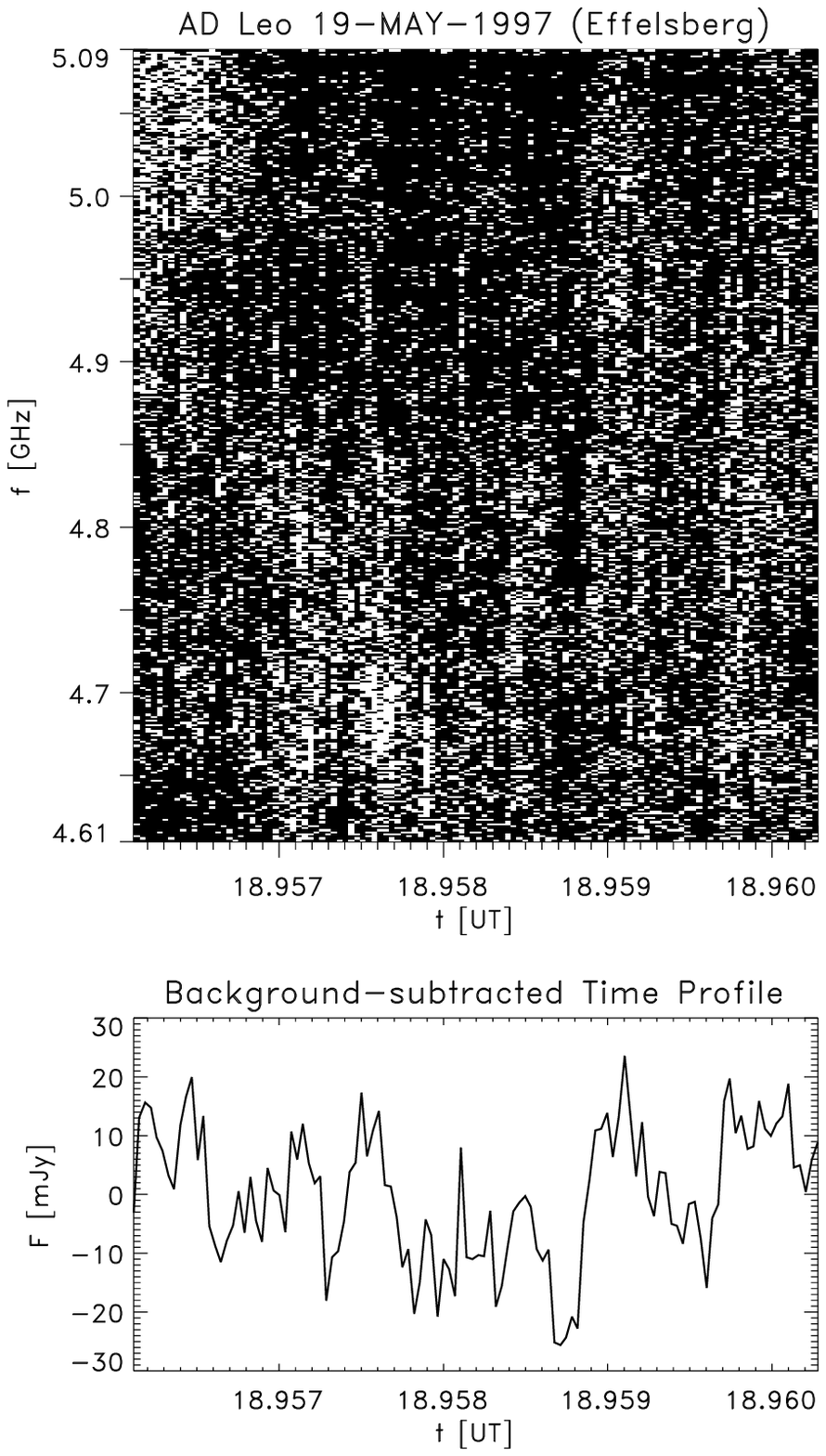}}
\caption[]
{Background-subtracted dynamic spectrum and time profile of the pulsating
structure contained in the burst, obtained by integrating the subtracted
dynamic spectrum over frequency.}
\label{radio-pulses}
\end{figure}

The burst showed an irregular pulsating structure during its decay phase
(18.956--18.961~UT), visible in the dynamic spectrum
(Fig.\,\ref{radio-spec} top left) as well as in the broad-band time
profiles (Fig.\,\ref{radio-spec} bottom).  This structure is shown in
greater detail in Fig.\,\ref{radio-pulses}.  In this figure, we have
subtracted the underlying flux level, obtained by taking the average over
the whole band of the spectral data set at 18.956~UT and 18.9602~UT and
linearly interpolating in between.  An irregular sequence of about 6
pulses is clearly discernible in the subtracted time profile shown in the
bottom panel.  The subtracted spectrum (Fig.\,\ref{radio-pulses}, top)
reveals also an irregular spectral structure.  The first three pulses do
not extend over the whole observed band and definitely show the flux
maximum at different frequencies.  The final two pulses seem to extend
over the whole band, i.e., their bandwidth exceeds 10\,\%.

\begin{figure}   
   \resizebox{\hsize}{!}{\includegraphics{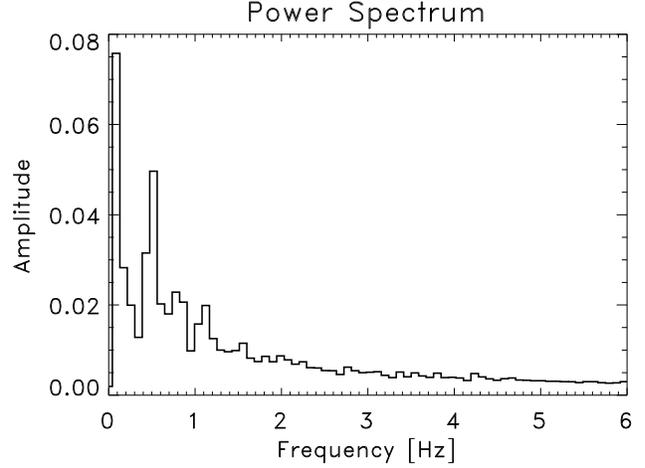}}
\caption[]
{Power spectrum of the time profile plotted in Fig.\,\ref{radio-pulses}.}
\label{power}
\end{figure}

The power spectrum of the pulse sequence (Fourier transform of the
autocorrelation function of the background-substracted time profile of
Fig.\,\ref{radio-pulses}) is shown in Fig.\,\ref{power}.  It reveals a
pulsation period of $\approx2$~s (0.5~Hz), which is clearly
different from the telescope oscillation period of $\approx1$~s. 
Furthermore, since the pulses possess spectral structure, they cannot be
of purely instrumental origin.

\section{Emission mechanism and source parameters}
\label{mechanism}

\subsection{Characteristics of M dwarf coronae}
\label{coronae}

First we recall a few basic findings about the physical conditions in the
atmospheres and coronae of M dwarf stars. Such stars possess outer 
convections zones and show universal X-ray emission 
(Schmitt \cite{Schm97}). Model fitting of low resolution 
X-ray spectra (Schmitt et al.\ \cite{Schm90}; Schmitt \cite{Schm97}) 
as well as studies of selected stars at significantly higher spectral 
resolution (Hawley et al.\ \cite{Haw95}) has shown that the coronae
of these stars contain plasma at a variety of temperatures with the emission
measure usually rising to beyond $T\ga2\times10^7$~K. Unfortunately,
direct measurements of density in the coronae of M dwarf stars
have not been made yet; Fossi \& Landini (\cite{FL94}) quote an
upper limit of $\sim3\times10^{12}$~cm$^{-3}$ for AU Mic. 
Katsova et al. (\cite{KBL87}) estimated the average density at the base of 
the corona from the X-ray emission measure of AD~Leo, assuming a
hydrostatic stratification and a temperature of $T=3\times10^6$~K, as
$n\sim5\times10^9$~cm$^{-3}$, 
a value probably smaller than the true value of density.
From kinetic pressure balance, one expects the hot component to possess 
a significantly smaller {\em average} density ($n\la10^9$~cm$^{-3}$) 
and because of the high temperature the corresponding scale height 
is of order of a stellar radius. The photospheric magnetic field strength
of AD~Leo is $B\approx3800$~G with a high filling factor of order 75\,\%
(Saar \& Linsky \cite{SL85}). The magnetic scale height is unknown; it is
often assumed to be smaller than the density scale height, but such
estimates are very model-dependent (see Saar \& Linsky \cite{SL85}).
Similar to the Sun, the plasma-beta of the corona must be very
small, even for $B\sim500$~G we find $\beta=4\times10^{-4}\ll1$
for the density and temperature given above. Hence, already a weak twist
in the magnetic field is able to support large perpendicular pressure
gradients, for example, loops with $T\ga2\times10^7$~K and much higher
densities of $n\sim5\times10^{11}$~cm$^{-3}$ embedded in an average 
corona. A filamentary corona, where magnetically confined dense loops, 
generally cooler than the average due to enhanced radiative 
or conductive losses, are embedded in
a dilute hot corona of lower field strength is consistent with the
observation (White et al.\ \cite{WLK94}) that the hot component in dMe
stars is permeated by relatively weak magnetic fields, $B<1000$~G.

\subsection{Estimate of the brightness temperature}
\label{estimate}

The brightness temperature of fully polarized radio emission is related to
the flux density $F_\nu$ by 
\begin{equation}
T_{\rm b}=\frac{c^2}{\kappa\nu^2}\,F_\nu\,\frac{d^2}{A}\,,
\label{T_b_0}
\end{equation}
where $\kappa$ is Boltzmann's constant, $d$ is the distance to the object,
and $A$ is the projected source area. 
Taking $A\sim\pi R_\ast^2$, one obtains $T_{\rm b}\sim5\times10^{10}$~K at
peak flux. This is a very conservative lower limit
on the brightness temperature, because the emission is expected to arise
from magnetically dominated structures, e.g., a loop or a group of loops,
which do not cover the whole stellar surface. 

By estimating the source extent from the timescale of the flux rise and a
characteristic velocity, one can attempt to obtain a tighter limit on
the brightness temperature. Let us first consider the velocity of the
radiating mildly relativistic electrons, $v_1$, which is appropriate if
the expansion of a cloud of fast electrons in a more or less stable
configuration, e.g.\ a magnetic trap, is seen as rise of the radio flux. 
The flare shows two main phases of flux rise. If extrapolated
backwards to zero flux level, both phases last $\sim\!7$~s. Using
$v_1\sim{c/3}$ for $\sim\!50$~keV particles, we obtain a source extent at
peak flux of $L_{\rm source}\la7\times10^{10}$~cm --- about two stellar
radii. As reviewed in Sect.\,\ref{coronae}, one cannot expect that the
high densities implied by the plasma emission mechanism or the high field
strengths implied by the ECM emission mechanism exist on such spatial
scales; rather, loops of much smaller height are to be expected.
Typically, loop heights are smaller than the pressure scale height (Rosner
et al.\ \cite{RTV78}), which is of order $\sim R_\ast/3$ for
$T\sim3\times10^6$~K. Therefore, the rise of the radio flux cannot be
explained as a source expansion with the velocity $v_1$. 

Alfv\'enic (or sub-Alfv\'enic) source expansion can be expected in case of
progression of the energy release process to neighbouring loops. Using 
$\omega_{\rm c}/\omega_{\rm p}\sim0.5$ (in accordance with the
interpretation of the burst as plasma emission), we obtain source sizes of
$A^{1/2}=L_{\rm source}\la2\times10^9$~cm and brightness temperatures of
$T_{\rm b}\ga3\times10^{13}$~K. 

Alternatively, the rise of the flux is due to an increasing number of fast
particles in the trap. For this case we suppose a loop length 
${\cal L}\sim0.1R_\ast$ based on the expectation that dense loops in
general are cooler and possess a smaller pressure scale height than the
average corona. The timescale of the flux pulsations, evident in
Figs.\,\ref{radio-pulses} and \ref{power}, provides an estimate of the
loop radius. If the pulsations are caused by radial MHD-oscillations of a
coronal loop (as suggested in Sect.\,\ref{puls}), their period is given by
$t_{\rm pulse}\sim r_{\rm loop}/v_{\rm A}$ (e.g., Aschwanden
\cite{Asch87}), which yields the loop radius as 
$r_{\rm loop}\sim7\times10^8$~cm. The resulting brightness temperatures
range from $T_{\rm b}\ga4\times10^{13}$~K (for 
$A\la2r_{\rm loop}{\cal L}$, taking into account that the trapped
particles develop a loss cone only in a section of the loop) to 
$T_{\rm b}\sim10^{14}$~K (for a source viewed along the loop axis,
$A\sim\pi r_{\rm loop}^2$). 

It should be noted that even higher $T_{\rm b}$ result in the latter case
if the source is filamentary at scales $\ll r_{\rm loop}$ (i.e., if
``loops'' are composed of fine ``threads''), as recent EUV observations of
the solar corona suggest (Reale \& Peres \cite{RP00}). Higher brightness
temperatures would also be obtained for Alfv\'enic source expansion if the
expansion is only one-dimensional (e.g., along an arcade of loops as in
large solar flares) or under the assumption of ECM emission, which
requires $\omega_{\rm c}/\omega_{\rm p}\ga1$. Although there remains a
substantial uncertainty in the value of the brightness temperature, all
estimates imply that the radiation is coherent. 

Two coherent mechanisms are generally considered relevant in the context
of stellar (and solar) radio emission: electron cyclotron maser (ECM)
emission and plasma emission. Both mechanisms are intrinsically
narrow-banded and require an inhomogeneous source to explain broadband
spectra. The inhomogeneity arises naturally in a magnetic trap, usually
supposed to be a magnetic loop, where energetic particles develop a
loss-cone distribution function (e.g., Dulk \cite{Dul85}). As discussed in
detail by Bastian et al.\ (\cite{BBDD90}), both mechanisms do have the
potential to explain a high brightness temperature, a high degree of
circular polarization, and strong time variability. However, the
absorption problem presents a severe difficulty for both mechanisms,
particularly for the ECM mechanism.

\subsection{Electron cyclotron maser emission}

For various models of a loss-cone distribution function of the energetic
particles and a wide range of background plasma temperatures
[$T\sim(0.2\mbox{--}1)\times10^7$~K] it was found that one can expect
growth of the electromagnetic modes due to the ECM instability only at the
first cyclotron harmonic, slightly above $\omega_{\rm c}$ (Sharma \&
Vlahos \cite{SV84}; Melrose et al.\ \cite{MHD84}). The extraordinary
($x$)~mode grows fastest in the range 
$\omega_{\rm p}/\omega_{\rm c}\la0.4$ and the ordinary ($o$)~mode grows
fastest in the range $0.4\la\omega_{\rm p}/\omega_{\rm c}\la1$. Maximum
growth occurs at large propagation angles, $\theta_{\rm m}\sim70^\circ$,
for both modes. Second and higher harmonics dominate the ECM instability
for $\omega_{\rm p}/\omega_{\rm c}\ga1$, but these waves cannot grow in
reality since the electrostatic and whistler mode instabilities attain
even higher growth rates in this range and exhaust the free energy
contained in the fast particle distribution (Sharma \& Vlahos \cite{SV84};
Winglee et al.\ \cite{WDP88}). (It should be noted that a somewhat
different result -- dominant growth of the $x$ mode at the second harmonic
for $\omega_{\rm p}/\omega_{\rm c}\ga0.6$ -- was obtained for {\em cold}
background plasma [Winglee \cite{Win85}]). Furthermore, the ECM
instability of mildly relativistic ($\sim30$~keV) electrons at the second
and higher harmonics is too weak to survive collisional damping for
background temperatures $T\ga1.5\times10^7$~K (supposing a density ratio
of accelerated electrons [$n_1$] to background electrons [$n$] of
$n_1/n\le10^{-3}$; see Fig.~3 in Sharma \& Vlahos \cite{SV84}). 

\begin{figure}  
   \resizebox{0.92\hsize}{!}{\includegraphics{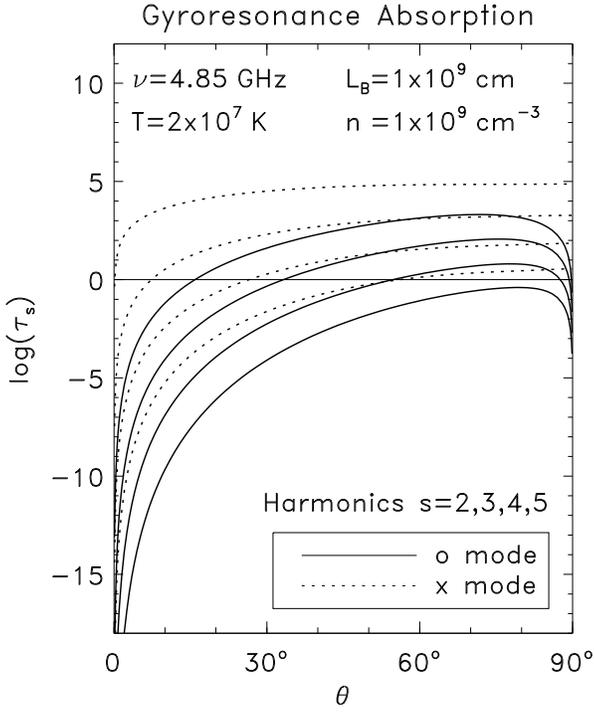}}
\caption[]
{Optical depths $\tau_s$ for gyroabsorption of microwave emission of
frequency $\nu=\omega/2\pi=4.85$~GHz at different harmonics $s$ vs.\ angle
between the magnetic field and the wave vector of the ordinary and
extraordinary mode for a Maxwellian plasma with $T=2\times10^7$~K,
$n=10^9$~cm$^{-3}$, and a magnetic field scale height
$L_B=10^9$~cm. The uppermost lines refer to $s=2$, and higher
harmonics show monotonically decreasing optical depth.}
\label{tau_ox}
\end{figure}

The modes that grow at the first harmonic are strongly damped by thermal
electrons as their ray path crosses the second, third, and fourth
cyclotron harmonic layers for decreasing magnetic field strength outside
of the source. This gyroresonance absorption increases with increasing
background plasma temperature and prevents propagation of the primary
growing ECM emission through the low harmonic layers in hot coronae.
Detailed calculations (Stepanov et al.\ \cite{Ste95, Ste99}) for realistic
coronal temperatures [$T\sim(0.3\mbox{--}3)\times10^7$~K] and densities
($n\sim10^9\mbox{--}10^{11}$~cm$^{-3}$) have shown that there is no escape
window for the $x$~mode at the second harmonic ($s=2$), only a very narrow
escape window ($\theta\la10^\circ$) for the $x$~mode at $s=3$ and the
$o$~mode at $s=2$, and only a moderately narrow escape window
($\theta\la30^\circ$) for the $o$~mode at $s=3$; also the fourth harmonic
contributes significantly to the damping. This is illustrated in
Fig.\,\ref{tau_ox} for parameters relevant to the burst under
consideration (where we have assumed that the magnetic field gradient
scale, $L_B\sim10^9$~cm, lies between the plausible values of the loop
radius and the loop length discussed in Sect.\,\ref{estimate}). The optical
depth scales roughly as $\tau_{o,x}\propto L_Bn/\omega$. A narrow escape
window of the $o$~mode at $s=3$ and nearly perpendicular propagation is
often discussed, but does not exist for our parameters and probably also
not for conditions of AD~Leo in general (note that in the figure we have
taken the minimum temperature of the hot coronal component and a rather
small value of $L_B$). 

ECM emission can reach the escape window at near-parallel propagation only
through a substantial amount of angle scattering. Such scattering, which
tends to isotropize the primary radiation pattern, is possible (Stepanov
et al.\ \cite{Ste99}), but its efficiency is difficult to estimate.
Considering the narrow angular interval of the ECM instability and the
narrow escape window, a reduction of the brightness temperature by about
one order of magnitude results from geometrical considerations alone.
Non-perfect efficiency of angle scattering is expected to lead to a
further significant reduction of the brightness temperature.

Assuming that nearly all of the energy contained in the fast
particles can be transferred to the maser emission, maximum intrinsic
brightness temperatures of ECM radiation at $\sim5$~GHz of $\sim10^{19}$~K
are obtained (Melrose \& Dulk \cite{MD84}). The saturation levels of the
ECM instability obtained from quasilinear theory (Aschwanden
\cite{Asch90}) or from the condition of phase locking between the
particles and waves (Wentzel \& Aschwanden \cite{WA91}) are strongly
dependent on the density and energy of the energetic particles but reach
values of $\sim10^{17}$~K or higher for typical parameters. With these
values, only a moderate efficiency of angle scattering
($10^{-3}\dots10^{-2}$) is required for ECM emission at the estimated
brightness temperature of $\sim2\times10^{14}$~K (for 
$\omega_{\rm c}/\omega_{\rm p}\sim2$). Hence, one cannot exclude
first-harmonic ($s\!=\!1$) $o$-mode ECM emission, but the difficulties with
the gyroresonance absorption do not suggest it to be the prime candidate
as was often assumed in the literature.

\subsection{Plasma emission}
\label{plasmarad}

Due to the strong gyroresonance absorption of ECM, we consider only the
plasma emission mechanism in the following, showing that it can, in fact,
explain the observations in spite of the high observing frequency. For a
trapped energetic particle distribution, plasma emission is based on the
electrostatic upper-hybrid instability. This instability was investigated
over a broad parameter range and for several slightly different loss-cone
distribution functions of the fast particles by Zheleznyakov \& Zlotnik
(\cite{ZZl75}), Winglee \& Dulk (\cite{WD86}), Stepanov et al.\
(\cite{Ste99}), and many others (see the references therein). Only the
range $\omega_{\rm p}>\omega_{\rm c}$ is relevant, since in the opposite
case, the growth rate of the electrostatic loss-cone instability drops
significantly and the gyroresonance absorption comes into play. The
electrostatic waves are always excited close to the upper-hybrid frequency
based on the density of the background plasma,
$\omega_{\rm uh}=(\omega_{\rm p}^2+\omega_{\rm c}^2)^{1/2}$, at
wavenumbers scaling with the cyclotron radius of the energetic particles,
$kv_1/\omega_{\rm c}\ga\omega_{\rm p}/\omega_{\rm c}$, and at
near-perpendicular propagation angles, $\theta_{\rm m}\ga80^\circ$.
Nonlinear scattering processes lead to radiation at the fundamental,
$\omega_{\rm f}\approx\omega_{\rm uh}$, and at the harmonic, 
$\omega_{\rm h}\approx2\omega_{\rm uh}$ (see Sect.\,\ref{bright}). We have
$\omega_{\rm uh}\sim\omega_{\rm p}$ for all relevant values of the
parameter $\omega_{\rm p}/\omega_{\rm c}$. 

If thermal effects are taken into account, the electrostatic waves and the
fundamental emission occur at 
$\omega_{\rm f}\approx\omega_{\rm uh}+3k^2v_T^2/2\omega_{\rm uh}$, where
$k$ is the wavenumber and $v_T=(\kappa T/m)^{1/2}$ is the electron thermal
velocity. This frequency lies below the $x$~mode cutoff frequency for 
$\omega_{\rm p}/\omega_{\rm c}\la20$ (e.g., Bastian et al.\
\cite{BBDD90}), which naturally leads to complete $o$~mode polarization.
Also harmonic plasma emission (which is much less affected by
gyroresonance and free-free absorption than fundamental plasma emission)
can lead to strongly polarized radio waves from hot coronae in the range
$1<(\omega_{\rm p}/\omega_{\rm c})^2\la5$, but polarization degrees of
$\approx100$\,\% over a considerable bandwidth, as observed in the present
event, would be difficult to explain (Stepanov et al.\ \cite{Ste99}).
Therefore, fundamental plasma emission has to be considered.

\begin{figure}  
   \resizebox{0.92\hsize}{!}{\includegraphics{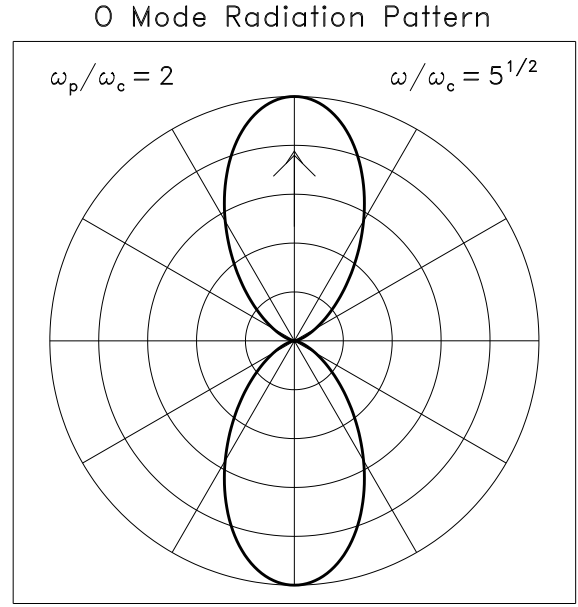}}
\vspace{0.3cm}
\caption[]
{Angular pattern of fundamental $o$ mode plasma emission resulting from
Rayleigh scattering of perpendicularly propagating plasma waves, according
to Eq.\,(\ref{anglepattern}). Radiated power is plotted at a linear scale.
The arrow indicates the direction of the magnetic field.}
\label{angle_f_o}
\end{figure}

Fundamental plasma emission at high frequencies suffers from free-free
absorption due to the implied high plasma densities. The free-free optical
thickness for fundamental plasma emission of frequency 
$\nu=\omega_{\rm p}/2\pi$ (neglecting for the moment the influence of the
magnetic field) is (Dulk \cite{Dul85}) 
\begin{equation}
\tau_{\rm ff}=1.5\times10^{-17}T^{-3/2}\nu^2L_n\,,
\label{tau_ff}
\end{equation}
where the density in the source is given by $\omega_{\rm p}$ and $L_n$ is
the scale length of the density decrease along the ray path. For
$\nu=5$~GHz ($n=3\times10^{11}$~cm$^{-3}$), the scale length leading to
$\tau_{\rm ff}=1$ becomes $\tilde{L}_n=2.7\times10^{-3}\,T^{3/2}$~cm. It
is immediately seen that the radiation could not escape if the radio
source were surrounded by plasma at $\sim3\times10^6$~K; the ray path must
run primarily through the hot coronal component. It is also clear that 
even the hot component of M dwarf coronae ($T<10^8$~K) would not permit
escape of the radiation if the implied high densities would slowly
decrease along the ray path according to a hydrostatic atmosphere:
$\tilde{L}_n(T\!=\!10^8~\mbox{K})\sim3\times10^9$~cm. However, as noted
above, such densities, being two orders of magnitude above the average,
can exist in the corona of AD~Leo only in thin filaments. If the external
density is $\le2\times10^9$~cm$^{-3}$, we obtain
$\tilde{L}_n(T\!=\!2\times10^7~\mbox{K})\ge3.5\times10^{10}~\mbox{cm}=
R_\ast$, permitting escape of the radiation. Since the average density of
the hot coronal component is lower than this value and since refraction
into regions of low density tends to ``guide'' the radiation through the
hot component, we expect that fundamental plasma emission at
$\nu\sim5$~GHz can escape. 

Whether free-free absorption is important also within the source remains
open. For $\nu=5$~GHz, $T=2\times10^7$~K, and 
$L_n=r_{\rm loop}\sim7\times10^8$~cm, we have $\tau_{\rm ff}\sim3$.
Under such conditions, the radiation can come from a surface layer of the
source only. However, $\tau_{\rm ff}<1$ for somewhat higher temperatures,
$T>4\times10^7$~K, which may exist even under quiescent conditions
(Schmitt et al.\ \cite{Schm90}) or may be exceeded due to the flare
heating. The calculations in the subsequent section suggest that the
free-free absorption within the source looses importance at sufficiently
high plasma wave energy densities (and brightness temperatures), where the
emission is stimulated. 

Fundamental plasma emission can be influenced also by gyroresonance
absorption. For $\omega_{\rm c}/\omega_{\rm p}\sim0.5$, the radiation has
to cross the harmonic layer $s=3$ and possibly even the layer $s=2$.
However, the effect is expected to be much weaker than for ECM emission 
in most cases, since fundamental plasma emission arises in the $o$ mode,
which is less damped than the $x$ mode, and in two symmetric cones
parallel and antiparallel to the magnetic field (Fig.\,\ref{angle_f_o}),
i.e., in the direction of the escape windows. Depending on the actual
value of $\omega_{\rm c}/\omega_{\rm p}$ and on changes of the magnetic
field direction outside of the inhomogeneous source, the emission may
escape without significant attenuation in some, perhaps many, cases and be
completely absorbed in others. 

Finally, we remark that, also due to the high coronal temperature, the
conversion efficiency of plasma waves into electromagnetic radiation in a
stellar environment exceeds the corresponding solar efficiency by one to
two orders of magnitude (Abada-Simon et al.\ \cite{Aba94a,Aba94b};
Stepanov et al.\ \cite{Ste99}).

\subsection{Estimate of source parameters}
\label{params}

From the condition $\omega_{\rm uh}/2\pi=4.85$~GHz, and using again
$\omega_{\rm c}/\omega_{\rm p}\sim0.5$ for the purpose of an estimate, we
obtain the plasma density $n\sim2.3\times10^{11}$~cm$^{-3}$ and the
magnetic field $B\sim770$~G. With $T=2\times10^7$~K, the plasma-beta
becomes $\beta\sim5\times10^{-2}$ and the mean free path is 
$\lambda_{\rm mfp}\sim10^7$~cm. With $T=5\times10^7$~K, we have
$\beta\sim0.13$ and $\lambda_{\rm mfp}\sim6\times10^7$~cm. For both
temperatures, the plasma is confined by the magnetic field and
$\lambda_{\rm mfp}\ll r_{\rm source}$, hence the source can in fact be
inhomogeneous, as is required for broadband plasma emission. The range of
densities and field strengths corresponding to the lower and upper limits
of the observed band is $n\sim(2.1\mbox{--}2.6)\times10^{11}$~cm$^{-3}$
and $B\sim730\mbox{--}810$~G. 

In comparison, if the burst were interpreted as ECM emission, we would
have $\omega_{\rm c}/2\pi\approx4.85$~GHz, equivalent to $B\approx1700$~G,
and the condition $\omega_{\rm p}<\omega_{\rm c}$ would lead to
$n<2.9\times10^{11}$~cm$^{-3}$ and $\beta<10^{-2}$. 

Minimum field strengths in the flare volumes of dMe stars have been
derived from the observed X-ray luminosities and the estimated source
volumes by van den Oord (\cite{vdO99}). Values $B\sim500\mbox{--}2000$~G
appear to be typical. Both ECM and plasma emission are consistent with
these estimates.

\section{Brightness temperature of plasma radiation}
\label{bright}

In this section we consider the relative role of fundamental (subscript f)
and harmonic (subscript h) plasma radiation from a flaring loop on AD~Leo.
The plasma waves are excited near the upper hybrid frequency, which is
expected to be rather close to the plasma frequency since 
$\omega_{\rm c}<\omega_{\rm p}$. They are transformed into electromagnetic
radiation at the fundamental 
($\omega_{\rm f}\approx\omega_{\rm uh}\approx\omega_{\rm p}$) by the
scattering on ions of the background plasma (Rayleigh scattering). The
conservation law for this scattering has the form
\begin{equation}
\omega_{\rm f}-\omega=(\vec{k}_{\rm f}-\vec{k})\cdot\vec{v}_i\,,
\label{Rayleigh}
\end{equation}
where $\omega_{\rm f}$, $\omega$ and $\vec{k}_{\rm f}$, $\vec{k}$ are the 
frequency and wave vector of electromagnetic and plasma waves,
respectively, and $\vec{v}_i$ is the particle (ion) velocity.

The second harmonic plasma radiation arises from a nonlinear coupling of
two plasma waves (Raman scattering), which obeys the resonance conditions
\begin{eqnarray}
\omega +\omega_1  &=& \omega_{\rm h}\,,                                  \\
\vec{k}+\vec{k}_1 &=& \vec{k}_{\rm h}\,. 
\label{k_res}
\end{eqnarray}

The magnetic field is important in determining the frequency spectrum of
the radiation which results from the conversion of plasma waves into
electromagnetic waves, but it does not affect the total power of the
radiation (Akhiezer et al.\ \cite{Akh75}). Moreover, for the loss-cone
instability, the spectrum of plasma waves can be considered as isotropic
(Zaitsev \& Stepanov \cite{ZS83}; Stepanov et al.\ \cite{Ste99}). Hence,
to estimate the brightness temperature, we can use the formulas for the
weak magnetic field case given by Zheleznyakov (\cite{Zhe96}) (see also
Zaitsev \& Stepanov \cite{ZS83}).

The solution of the transfer equation for the brightness temperature of 
radiation can be represented in the form 
\begin{equation}
T_{\rm b}=\frac{\alpha}{\mu_{\rm c} + \mu_{\rm nl}}\, 
          \left\{1-\exp\left[-\!\int(\mu_{\rm c}+\mu_{\rm nl})\,{\rm d}l
	               \right]\right\}  
\label{transfer}
\end{equation}                     
with the emissivities for the fundamental and the second harmonic 
\begin{eqnarray}
\alpha_{\rm f}&\approx&\frac{\pi}{36}\,\frac{\omega_{\rm p}}{v_{\rm g}}\, 
                       \frac{m v_1^2}{\kappa}\, w \,,                    \\
\alpha_{\rm h}&\approx&\frac{(2\pi)^5}{15\sqrt{3}}\,
                     \frac{c^3}{\omega_{\rm p}^2v_1}\,\frac{w^2}{\xi^2}\,nT
\end{eqnarray}                
and the absorption coefficients due to the electron-ion collisions
\begin{equation}
\mu_{\rm c}=\omega_{\rm p}^2\nu_{\rm ei}/\omega^2 v_{\rm g}\,,
\end{equation}
due to the scattering of the waves on background plasma ions 
\begin{equation}
\mu_{\rm nlf} \approx 
  -\frac{\pi}{108}\,\frac{m}{M}\,\frac{\omega_{\rm p}^3}{v_{\rm g}}\,
   \frac{1}{nTv_T^2}\,\frac{1}{k}\,\frac{\partial}{\partial k}\,(kW_k)\,,
\end{equation}
and due to the decay of an electromagnetic wave into two plasma waves 
\begin{equation}
\mu_{\rm nlh} \approx 
  \frac{(2\pi)^2}{15\sqrt{3}}\,\frac{\omega_{\rm p}}{v_1}\,\frac{w}{\xi}\,.
\end{equation}
Here 
\begin{equation}
\nu_{\rm ei}=5.5nT^{-3/2}\ln(10^4T^{3/2}n^{-1/3})
\end{equation}
is the electron-ion collision frequency, 
$v_{\rm g}=\sqrt{3}k(l)v_Tc/\omega_{\rm p}$ and $v_{\rm g}=\sqrt{3}c/2$
are the group velocities of the electromagnetic waves at the fundamental
and at the second harmonic, respectively, $v_T=(\kappa T/m)^{1/2}$ is the
thermal velocity of the background electron component,
$v_1=c[1-(mc^2/(\kappa T_1+mc^2))^2]^{1/2}$ is the velocity of the
energetic electrons, $m/M$ is the mass ratio, and $w=W/n\kappa T$ is the
normalized energy density of the electrostatic waves, where 
$W=\int W_k{\rm d}k$. The formal parameter $\xi$ is related to the width
of the electrostatic wave spectrum by 
\begin{equation} 
(\Delta k)^3\sim\frac{4\pi}{3}\,(k_{\max}^3-k_{\min}^3)
            =   \xi(\omega_{\rm p}/c)^3\,. 
\end{equation}
We shall use $k_{\min}\sim\omega_{\rm p}/c$ and 
$k_{\max}\sim\omega_{\rm p}/5v_T$ (which takes damping of the
electrostatic waves at the thermal background into account and implies
$T_1\ga25T$). 

For dominance of fundamental emission, we require stimulated emission
(maser effect), i.e., a negative absorption coefficient. This happens for
$\partial(kW_k)/\partial k>0$, which is valid if the plasma wave spectrum
is sufficiently flat or has a positive slope. This condition is easily
satisfied for plasma wave turbulence spectra formed by nonlinear processes
(Kaplan \& Tsytovich \cite{KT73}). 

The exponent in Eq.\,(\ref{transfer}) is evaluated as 
$\tau=\int_0^L\mu{\rm d}l$, assuming a flat plasma wave spectrum. Since
the density in the source is inhomogeneous with a scale length
$L_n=|n/\nabla n|$, the integral is carried out only in a thin layer of
depth 
\begin{equation} 
L \approx 3L_n\,\frac{v_T^2}{\omega_{\rm p}^2}(k_{\max}^2-k_{\min}^2)\,, 
\end{equation} 
in which the frequency of the electromagnetic waves is approximately
constant. Since $k_{\max}^2\ll\omega_{\rm p}^2/v_T^2$, we have $L\ll L_n$.
Furthermore, the relation ${\rm d}l=6L_nv_T^2k{\rm d}k/\omega_{\rm p}^2$
is used (see Zai\-tsev \& Stepanov [\cite{ZS83}] for details). The
resulting brightness temperatures for fundamental and harmonic radiation
are 
\begin{eqnarray}
T_{\rm bf}&=&\frac{Aw}{\nu_{\rm ei}-Cw}
             \left(1-e^{-B(\nu_{\rm ei}-Cw)L_n}\right)\,,    \label{T_bf}\\
T_{\rm bh}&=&\frac{Dw^2/\xi^2}{\nu_{\rm ei}+Fw/\xi}
             \left(1-e^{-E(\nu_{\rm ei}+Fw/\xi)L}\right)\,,
	                                                     \label{T_bh}
\end{eqnarray}
where we have introduced the abbreviations 
\begin{eqnarray}
A&=&\frac{\pi}{36}\frac{v_1^2}{v_T^2}\omega_{\rm p}T\,,                  \\
B&=&2\sqrt{3}\,\frac{v_T}{c}(k_{\rm max}-k_{\rm min})\omega_{\rm p}^{-1}\,,\\
C&=&\frac{\pi}{108}\frac{m}{M}\frac{v_1^2}{v_T^2}\omega_{\rm p}\,,       \\
D&=&\frac{2(2\pi)^5}{15}\frac{nc^4}{\omega_{\rm p}^2v_1}T\,,             \\
E&=&(2\sqrt{3}\,c)^{-1}\,,                                               \\ 
F&=&\frac{2(2\pi)^2}{15}\frac{c}{v_1}\omega_{\rm p}\,.
\end{eqnarray}

\begin{figure}[!t]  
   \resizebox{\hsize}{!}{\includegraphics{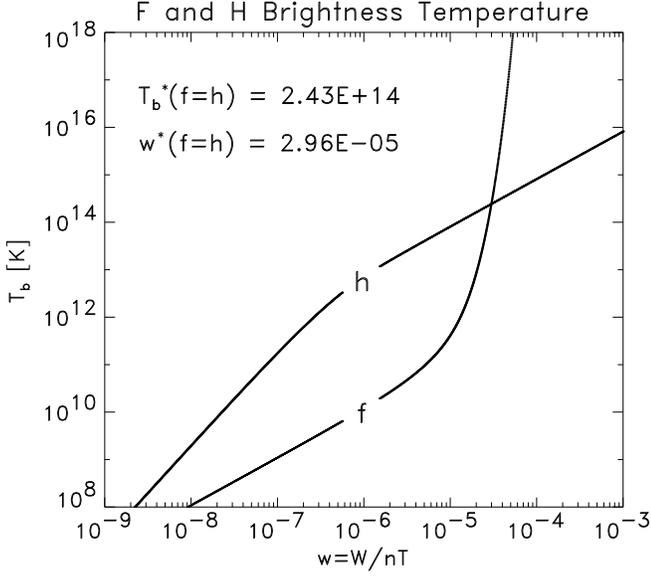}}
\caption[]
{Fundamental and harmonic brightness temperature vs.\ plasma turbulence 
level according to Eqs.\,(\protect\ref{T_bf}) and (\protect\ref{T_bh})
for $\omega_{\rm p}/2\pi=4.85$~GHz, $T=2\times10^7$~K, 
$T_1=5\times10^8$~K (43~keV), and $L_n=10^9$~cm.}
\label{turb-level}
\end{figure}

The brightness temperatures are plotted in Fig.\,\ref{turb-level} for
plasma parameters supposed to be representative of a flaring loop on
AD~Leo ($T=2\times10^7$~K, $T_1=5\times10^8$~K, $L_n=10^9$~cm), i.e., we
suppose that the dense loop is heated up by the flare to about the
temperature of the hot coronal component. The thermal level of plasma wave
fluctuations for these parameters is 
$w_{\rm th}\sim(6\pi^2nr_{\rm D}^3)^{-1}\approx8\times10^{-10}$ (where
$r_{\rm D}=v_T/\omega_{\rm p}$ is the Debye radius). An exponential
increase of the brightness temperature of fundamental radiation
(stimulated emission) occurs for $w\ga3\times10^{-6}$. The two curves
intersect at $w^\ast\approx3\times10^{-5}$, and 
$T_{\rm b}^\ast\approx2\times10^{14}$~K. Fundamental emission dominates
above this level of plasma turbulence. Our estimates in
Sect.\,\ref{estimate} suggested a brightness temperature 
$T_{\rm b}\ga3\times10^{13}$~K, which is roughly in accordance with the
calculated minimum fundamental brightness temperature, given the
uncertainty of the estimate and the
approximate nature of Eqs.\,(\ref{T_bf}) and (\ref{T_bh}). 

Fundamental radiation reaches brightness temperatures exceeding the
estimated lower limit of $\sim5\times10^{10}$~K also in the range
$3\times10^{-6}\la w<w^\ast$, in which a brighter continuum around
10~GHz due to harmonic plasma radiation would have remained unobserved.
Such a situation cannot be excluded but does not appear very probable,
since it is not supported by solar radio observations. 

The brightness temperature of fundamental radiation increases rapidly with
increasing $w$, reaching $T_{\rm bf}>10^{20}$~K within a short interval of
$w$ values. Strong turbulence effects will then limit the brightness
temperature to $T_{\rm bf}<<10^{22}$~K, at which level the energy density
in the electromagnetic waves becomes comparable to the thermal
(background) plasma energy density. Viewed reversely, in the regime of
dominating fundamental emission, the plasma wave energy density $w$ stays
within a rather small range. 

We have to ask whether the plasma wave energy density at saturation of the
electrostatic upper-hybrid instability of the trapped particles can reach
the level $w\sim3\times10^{-5}$. A summary of the diverging answers to
this question, found in the literature, was given by Stepanov et al.\
(\cite{Ste99}). We mention here only the estimates quoted by Dulk
(\cite{Dul85}), $w\sim10^{-5}$, which are consistent with
our interpretation of the burst as fundamental plasma emission, and the
estimate by Zaitsev et al.\ (\cite{Zai97}), who considered the case of
saturation by moderate quasilinear diffusion (with almost isotropic
energetic electrons): 
\begin{equation}
w \approx \frac{n_1}{n}\,\frac{T_1}{T}\,\frac{\ln\sigma}{\sigma}\,
          \frac{v_1}{\nu_{\rm ei}{\cal L}}\,.
\label{w_level}
\end{equation}
Here $\sigma$ is the mirror ratio of the trap and $\cal L$ is the loop
length. Taking the same parameters as in Fig.\,\ref{turb-level} with
$\sigma\sim3$ and ${\cal L}\sim3\times10^9$~cm, the latter estimate yields
$w\sim0.23\,n_1/n$. Dominance of fundamental plasma emission then requires a 
density ratio $n_1/n\sim10^{-4}$, 
which is quite high but still reasonable (it can be reduced somewhat if
$T_1/T$ is larger). We conclude that the 4.85~GHz emission in the
considered radio flare can be due to fundamental plasma radiation. 

A comparison with the case of decimetric emission is of interest.
Repeating the calculation for a density that corresponds to a plasma
frequency $\omega_{\rm p}/2\pi=1.4$~GHz with the same values of $T$, $T_1$,
and $L_n$ gives the crossing point of the fundamental and harmonic
brightness temperatures at $w^\ast\sim10^{-4}$ and 
$T_{\rm b}^\ast\sim3\times10^{15}$~K. Hence, harmonic plasma emission
tends to dominate at longer wavelengths (Stepanov et al.\ \cite{Ste99}). 

\begin{figure}[!t]   
   \resizebox{\hsize}{!}{\includegraphics{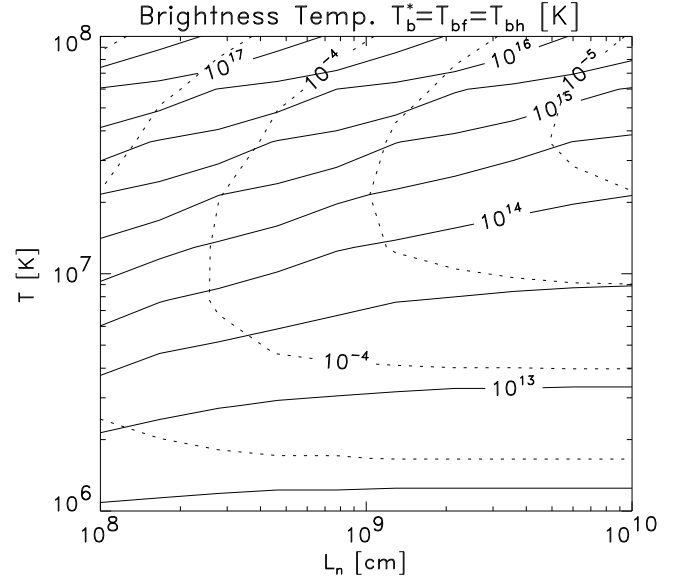}}
\caption[]
{Brightness temperature (solid) and plasma wave level (dashed) at the 
crossing point, $T_{\rm bf}(w^\ast)=T_{\rm bh}(w^\ast)$ 
(cf. Fig.\,\ref{turb-level}), in $T$--$L_n$ parameter space for 
$\omega_{\rm p}/2\pi=4.85$~GHz and $T_1=25T$. The contour lines are a 
factor $10^{1/2}$ apart.}
\label{param}
\end{figure}

An overview of the parameter dependence of the brightness temperature and
plasma wave energy density at the intersection 
$T_{\rm bf}(w)=T_{\rm bh}(w)$ is given in Fig.\,\ref{param}. It is seen
that this brightness temperature increases with increasing background
temperature and decreasing density gradient scale. Brightness temperatures
$T_{\rm b}>10^{13}$~K are achieved in a large range of parameter space.

\section{Electron acceleration}
\label{accel}

In order to provide long-lasting radio radiation from a coronal loop,
emitted by nonthermal electrons, a permanent source of energetic electrons
is required, because the particles are lost from the magnetic trap due to
diffusion into the loss cone. Under steady-state conditions, the
acceleration rate $\dot{N}$ equals the loss rate into the loss cone. The
particle loss rate due to pitch-angle diffusion of electrons by scattering
on waves, which are excited by the loss-cone instability, is of the order
$2n_{\rm prec}v_1S$ for two-sided precipitation in a nearly symmetrical
loop, where $n_{\rm prec}$ is the average number density of precipitating
electrons and $S$ is the loop cross-sectional area at the footpoints. The
fraction of precipitating electrons was estimated by Stepanov \& Tsap
(\cite{ST99}) as $n_{\rm prec}\sim n_1/(2\sigma)$. Hence,
\begin{equation}
\dot{N} = n_1v_1S/\sigma\,.
\label{N_dot1}
\end{equation}
Combining Eqs.\,(\ref{w_level}) and (\ref{N_dot1}), we find 
\begin{equation}
\dot{N} = \frac{n\nu_{\rm ei}{\cal L}S}{\ln \sigma}\,\frac{T}{T_1}\,w\,.
\label{N_dot2}
\end{equation}
The level of plasma turbulence, $w$, can be estimated from the
calculations of Sect.\,\ref{bright}. Figure\,\ref{turb-level} shows that
$w$ stays rather close to $w^\ast$ in the range where
fundamental plasma emission dominates. 

We assume plasma parameters in the coronal loop as follows: 
$n=2.3\times10^{11}$~cm$^{-3}$, $T=2\times10^7$~K, $T_1=5\times10^8$~K
(43~keV), ${\cal L}=3\times10^9$~cm, $S\sim10^{18}$~cm$^2$, $\sigma=3$.
From Fig.\,\ref{turb-level} we then have $w^\ast\sim3\times10^{-5}$, and
Eq.\,(\ref{N_dot2}) yields the acceleration rate required for the observed
continuum brightness temperature as 
$\dot{N}\sim10^{35}$ electrons~s$^{-1}$, which is nearly of the same order
as the electron acceleration rate during large solar flares (e.g., Miller
et al.\ \cite{Mil97}). 

One possible origin for the particle acceleration in
coronal magnetic loops on red dwarf stars is the production of run-away
electrons in sub-Dreicer quasi-stationary electric fields, which can be a
very efficient acceleration mechanism. Most
authors who considered electron acceleration in DC electric fields have
used the electric field of the currents in a plasma with the Spitzer
resistivity or with anomalous resistivity (e.g., Holman \cite{Hol85};
Tsuneta \cite{Tsu85}). 
Zaitsev et al.\ (\cite{ZUS00}) considered another situation, namely the
electron acceleration by electric fields that result from charge
separation, which is caused (for strongly different electron and ion
magnetization) by flows of partly ionized plasma at the loop footpoints.
They obtained the electric field component parallel to the magnetic field
as 
\begin{equation}
E_\parallel = \frac{1-{\cal F}}{{\cal F}^2}\,\frac{M\nu_{\rm in}v_r}{e}\,
              \frac{B_r}{B}\,,
\end{equation}
where ${\cal F}=n_{\rm n}m_{\rm n}/(n_{\rm n}m_{\rm n}+nM)$ is the
relative density of neutrals, $\nu_{\rm in}$ is the ``effective''
frequency of ion-neutral collisions, $B_r$ is the small radial component
of the magnetic field in a flux tube arising due to the expansion of the
flux tube from the photosphere to the chromosphere, and $v_r$ is the
radial component of the flow velocity. For sub-Dreicer fields,
$E_\parallel<E_{\rm D}=e\,r_{\rm D}^{-2}\ln\Lambda$, the electron
acceleration rate is
\begin{equation}
\dot{N}=0.35\,n_{\rm a}\nu^{\rm a}_{\rm ei}V_{\rm a}
        \left(\frac{E_{\rm D}}{E_\parallel}\right)^{3/8}
	\exp\left[-\sqrt{\frac{2E_{\rm D}}{E_\parallel}}
	          -\frac{E_{\rm D}}{4E_\parallel}\right]
\label{N_dot3} 
\end{equation}
(Knoepfel \& Spong \cite{KS79}). Here the index ``a'' denotes quantities
in the acceleration region, which has a volume $V_{\rm a}$. 

To estimate the resulting acceleration rate, we
take the parameters of the acceleration region in the stellar chromosphere
as follows: $T_{\rm a}=10^5$~K, $n_{\rm a}=10^{12}$~cm$^{-3}$, 
${\cal F}\approx10^{-2}$, and $v_r\approx5\times10^4$~cm\,s$^{-1}$
(similar to fast solar convection). With $B_r\sim0.1~B$ we obtain
$E_\parallel\approx10^{-2}$~V\,cm$^{-1}\approx E_{\rm D}/30$. An
acceleration length $h=100$~km is then sufficient for the acceleration of
$\sim100$~keV electrons. To obtain an acceleration rate of
$\dot{N}\sim10^{35}$~s$^{-1}$, the cross section of the chromospheric
acceleration volume has to be $\sim10^{16}$~cm$^2$, which is consistent
with the chosen value of $B_r$ and the coronal loop radius estimated in
Sect.\,\ref{estimate}. The density of energetic electrons required for
the observed radio fux was estimated in Sect.\,\ref{bright} as
$n_1\sim10^{-4}n\sim2\times10^7$~cm$^{-3}$. This can be supplied by the
chromospheric accelerator to a coronal loop with a volume of
$10^{27}\mbox{--}10^{28}$~cm$^3$ in $\sim\!1$~s. We conclude that this
process is able to explain the acceleration underlying the
observed intense radio burst.

\section{Origin of pulsations}
\label{puls}

There are indications of flux pulsations with an average period of
$\approx2$~s during the decay phase of the  event under consideration
(Figs.\,\ref{radio-pulses}, \ref{power}). Usually four modulation
mechanisms for stellar and solar radio emission are discussed in the
literature:
(1) modulation due to MHD oscillations of a flux tube containing the
source, 
(2) intensity oscillations of the underlying electrostatic waves during
the nonlinear stage of the wave-wave or wave-particle interactions, 
(3) modulation of the particle acceleration rate, and 
(4) modulation due to electric current oscillations in an equivalent 
LRC-circuit (Aschwanden \cite{Asch87}; Zaitsev et al.\ \cite{Zai98}).

In order to obtain estimates of the radio source dimension, we considered
the first mechanism of pulsations in Sect.\,\ref{estimate}, supposing
radial oscillations of a loop which have a period 
$t_{\rm pulse}\sim r_{\rm loop}/v_{\rm A}$, and obtained a quite
reasonable value for the loop radius, $r_{\rm loop}\sim7\times10^8$~cm. 

As for mechanism (2), the maximum value of the pulsation period can be
estimated as $t_{\rm nl}\approx2\pi/\nu_{ei}\sim10^{-2}$~s (Zaitsev
\cite{Zai71}; Aschwanden \cite{Asch87}), which is too small in comparison
to the observed period. 

The period of eigen-mode oscillations in a current-carrying coronal loop
modeled as an equivalent LRC-circuit is $t_{\rm LC}=(2\pi/c)(LC)^{1/2}$.
The total inductance of a slender loop can be estimated as
$L\approx10{\cal L}$ (Alfv\'en \& Carlquist \cite{AC67}). The capacitance
depends on the current $I$ flowing through the cross section of a loop,
because just this current determines the self-consistent magnetic field
and the effective dielectric permittivity of a loop plasma: 
$C(I)\approx c^4nMS^2/(2\pi{\cal L}I^2)$ (Zaitsev et al.\ \cite{Zai98}).
Supposing again $S\sim10^{18}$~cm$^2$, we obtain 
$t_{\rm LC}\sim3\times10^{13}/I$. Taking  $t_{\rm LC}\approx2$~s, we find
the value of the electric  current, $I\sim1.5\times10^{13}$~A, and thus
the magnetic field in the loop as $B\sim5$~kG. This magnetic field value
exceeds the photospheric field strength, which is highly improbable for
the 4.85~GHz source. 

Hence, only a modulation of the radio source by MHD oscillations or a
modulated particle acceleration appear to be consistent with the observed
period of the radio pulsations. It should be noted that MHD oscillations
of a loop can lead also to a quasi-periodical regime of electron
acceleration or electron injection into the loop. Alternative models
exist, for example the quasi-periodic modulation of particle acceleration
by dynamic magnetic reconnection (Kliem et al.\ \cite{KKB00}). However,
the fact that the pulsations are more clearly developed during the
decay phase of the flare suggests MHD oscillations of a loop, which can be
triggered by the primary flare energy release.

\section{Conclusions}
\label{concl}

\hspace{\parindent}\hspace{-0.25em}
1. Coherent plasma radiation can explain the intense broad-band AD~Leo
radio flare on 19 May 1997 at microwave frequencies (4.6--5.1~GHz), whose
brightness temperature is estimated to be at least of order 
$T_{\rm b}\sim5\times10^{10}$~K (with values 
$T_{\rm b}\ga3\times10^{13}$~K considered to be more probable). 
Electron cyclotron maser emission appears less probable than plasma emission,
primarily due to the strong gyroresonance absorption in the hot stellar
corona, but cannot be excluded. 

2. In case of the plasma emission mechanism, the circular polarization
degreee of $\approx100$\,\% requires emission at the fundamental plasma
frequency. Harmonic plasma radiation is less strongly polarized. 

3. The parameters of the radio source, assumed to be a magnetic loop, were
estimated for plasma emission as $n\sim2.3\times10^{11}$~cm$^{-3}$ and
$B\sim800$~G. Supposing $T\sim2\times10^7$~K, the plasma-beta is of order
$\beta\sim5\times10^{-2}$. 

4. A strongly inhomogeneous corona is required to avoid complete free-free
absorption of the plasma emission. The ratio of source and external
densities must be at least of order $n/n_{\rm ext}\sim10^2$. 

5. Acceleration of electrons by DC electric fields at the footpoints of a
loop, driven by photospheric convection, can give a rate of electron
acceleration, $\dot{N}\sim10^{35}$~s$^{-1}$, sufficient to explain the
observed flare in terms of plasma emission. 

6. A possible interpretation of the quasi-periodic modulation of the 
broadband radio emission during the decay phase of the flare is given by
radial MHD oscillations of a flaring loop, which yields the loop radius as
$r_{\rm loop}\sim7\times10^8$~cm.

\begin{acknowledgements}
We acknowledge the help by the operators of the Effelsberg 100~m
radiotelescope and thank Dr.\ J.\ Neidh\"ofer for his assistance in
setting up the observations.
We thank the referee for constructive comments. 
A.V.\,S.\ acknowledges an invitation to the Astrophysical Institute Potsdam.
A.V.\,S.\ and V.V.\,Z.\ were supported by RFBR grants No.\ 99-02-16045 and
00-02-16356, and by the program ``Astronomy'';
B.K.\ was supported by BMBF/DLR grant No.\ 01\,OC\,9706\,4, and
J.H.\ was supported by DFG grant No.\ STA351/6-1.
\end{acknowledgements}

\appendix

\section{Angular pattern of fundamental plasma emission}

The angular pattern of fundamental plasma emission resulting from Rayleigh
scattering of perpendicularly propagating plasma waves near the electron
plasma frequency was obtained in Stepanov (\cite{Ste70}) as
\begin{equation}
P(\theta) = \frac{A^2+B^2}{C}\,N^3
\label{anglepattern}
\end{equation}
with 
\begin{eqnarray}
A&=&\varepsilon_1-1+\frac{\varepsilon_2^2}{N^2-\varepsilon_1}\,,\nonumber\\
B&=&\varepsilon_2\,\frac{N^2-1}{N^2-\varepsilon_1}\,,           \nonumber\\
C&=&\varepsilon_1+
    \varepsilon_1\,\frac{\varepsilon^2_2}{(N^2-\varepsilon_1)^2}+
    \frac{2\,\varepsilon^2_2}{N^2-\varepsilon_1}+               \nonumber\\
 & &\varepsilon_3\,\frac{N^4\sin^2\theta\,\cos^2\theta}
                        {(N^2\sin^2\theta-\varepsilon_3)^2}\,.  \nonumber
\end{eqnarray}
The elements of the cold-plasma dispersion tensor, $\varepsilon_i$, and
the refractive index, $N$, are as given in Stepanov et al.\
(\cite{Ste99}). Note that the assignment of the $o$ and $x$~modes to the
sign of terms in the expression for $N$ was erroneously exchanged in that
paper.

\end{document}